\documentclass[%
reprint,%
amssymb, amsmath,%
aip,cha,%
]{revtex4-1}

\usepackage{docs}%
\usepackage{graphicx}  
\usepackage{bm}%
\usepackage[colorlinks=true,linkcolor=blue]{hyperref}%
\usepackage{soul}
\usepackage{color}
\usepackage[table]{xcolor}
\definecolor{BLACK}{gray}{0}
\newcommand{\blk}{\color{BLACK}}

\expandafter\ifx\csname package@font\endcsname\relax\else
\expandafter\expandafter
\expandafter\usepackage
\expandafter\expandafter
\expandafter{\csname package@font\endcsname}%
\fi
\begin{document}

\title{Photonic quantum information processing: a concise review}%
\author{Sergei Slussarenko}%
\email{s.slussarenko@griffith.edu.au}
\affiliation{Centre for Quantum Dynamics \& Centre for Quantum Computation and Communication Technology, Griffith University, Brisbane, QLD 4111, Australia.}%
\author{Geoff J. Pryde}%
\affiliation{Centre for Quantum Dynamics \& Centre for Quantum Computation and Communication Technology, Griffith University, Brisbane, QLD 4111, Australia.}%

\begin{abstract}
Photons have been a flagship system for studying quantum mechanics, advancing quantum information science, and developing quantum technologies.  Quantum entanglement, teleportation, quantum key distribution and early quantum computing demonstrations were pioneered in this technology because photons represent a naturally mobile and low-noise system with quantum-limited detection readily available. The quantum states of individual photons can be manipulated with very high precision using interferometry, an experimental staple that has been under continuous development since the 19th century. The complexity of photonic quantum computing device and protocol realizations has raced ahead as both underlying technologies and theoretical schemes have continued to develop. 
Today, photonic quantum computing represents an exciting path to medium- and large-scale processing. It promises to put aside its reputation for requiring excessive resource overheads due to inefficient two-qubit gates. Instead, the ability to generate large numbers of photons---and the development of integrated platforms, improved sources and detectors, novel noise-tolerant theoretical approaches, and more---have solidified it as a leading contender for both quantum information processing and quantum networking. Our concise review provides a flyover of some key aspects of the field, with a focus on experiment. Apart from being a short and accessible introduction, its many references to in-depth articles and longer specialist reviews serve as a launching point for deeper study of the field.
\end{abstract}

\maketitle

\tableofcontents

\section{Introduction}

\subsection{Optical quantum computing}

With the invention of the quantum computing (QC) concept, the development of suitable optical quantum technology became both an interesting approach to the problem, and a necessity. On one hand, the advantages of using photons as information carriers seem to be obvious: photons are clean and decoherence-free quantum systems for which single-qubit operations can be easily performed with incredibly high fidelity~\cite{peters03}. On the other hand, quantum information handling with photons as ``flying qubits'' is required for communication-based quantum information science tasks, such as networking quantum computers and enabling distributed processing. 

In terms of the traditional DiVincenzo criteria of a quantum computer~\cite{divincenzo00}, five out of seven are essentially satisfied by choosing photons. The remaining criteria are harder to satisfy because photons don't easily interact, making deterministic two-qubit gates a challenge. Among the additional technical considerations is photon loss, which arises from currently-imperfect detection and photon generation techniques, and from scattering and absorption in optical components comprising the computation circuits. And although photons are always flying, computing and networking tasks may need them to be delayed or stored, so an extra device---an optical quantum memory---may sometimes be needed. Addressing each of these considerations requires additional resources, creating a notionally large optical QC overhead that has sometimes led to negative perceptions of the photonic approach.
 
Of course, there is intense research underway in the development of deterministic optical (but matter-mediated) quantum gates~\cite{tiecke14,tiarks16,langford11}, which could take photonic quantum computing in a new direction. Meanwhile, the idea of linear optical quantum computing (LOQC) that relies on simple, but probabilistic, quantum operations has increasing promise as it has continued development over the last 20 years. The earlier history of the field is covered in previous reviews~\cite{rev_ralph09,rev_kok07,rev_obrien09,rev_ladd10} that have appeared regularly in the literature. Here, we do not provide a typical review---that is, we do not present a comprehensive encapsulation of all the achievements of the field during the past decade. Instead, we concentrate on the few technological, experimental and theoretical advances that we think play key roles on the path towards a universal quantum computer operating with individual photons and linear operations. On the technology side, we look at photon detection and generation tools, and integrated waveguide technology---and some new intermediate quantum computing demonstrations that these enable. On the conceptual side, we discuss a few promising ways towards a realistic universal linear optical quantum computer.  We will concentrate on photonic\footnote{In this article, we use the term `photonic' to refer to schemes with counted photons, i.e. to discrete-variable schemes such as those with qubit $\equiv$ photon or qudit $\equiv$ photon.} quantum computing (PQC) that relies on qubits encoded in discrete variables, noting, however, that quantum computing with continuous variables has now become an important part of LOQC~\cite{lloyd99,rev_braunstein05,menicucci06,lenzini18a}. But before that, we start with a brief refresher on the basic conceptual elements and history of PQC.

\subsection{Basics}
A qubit can be encoded as probability amplitudes corresponding to the photon occupation of two modes of some degree of freedom of the optical field. This method is known as dual-rail encoding.
The most commonly-used mode pairs are orthogonal polarizations or non-overlapping propagation paths, but recently, other degrees of freedom such as transverse spatial~\cite{dada11,dambrosio12,rev_erhard18}, frequency mode~\cite{roslund13,cai17,kues17,lukens17},  temporal bin-~\cite{brendel99,humphreys13,jayakumar14,samara19} and  temporal mode-~\cite{kielpinski11,humphreys13,brecht15,averchenko17,ansari18} encoding are attracting attention. One-qubit operations---i.e. the shifting of single-photon population between the two modes that comprise the dual-rail qubit, and applications of phase shifts between them---are easily and reliably implemented using interferometry in the degree of freedom of choice. A great advantage of optical quantum computing is that it does not have to be confined to qubits: many of the degrees of freedom listed provide a natural way to encode multi-level qudits. Moreover, several degrees of freedom of the same photon can be used simultaneously~\cite{gao10,graham15,wang15,malik16,wang18}. (As we will discuss later, these tools provide a natural advantage for optics, allowing for simpler logical circuits even when working with qubits as the basic logical elements.) A way to realize an arbitrary $n$-dimensional unitary transformation on the mode space, with linear optics, has been outlined by Reck \textit{et al}.~\cite{reck94} quite some time ago, with recent improvements~\cite{clements16} and expansions~\cite{tischler18}. In principle, Reck-type schemes could perform universal processing with a single photon in many modes used to represent multiple qubits. Unfortunately, that encoding leads to exponential scaling in the number optical components, and thus cannot be used to build a scalable quantum computer. Thus the use of multiple single photons is required for circuits with two-qubit gates and beyond.

It is natural, then, to implement one qubit per photon, with a dual-rail encoding. Two-qubit operations require the ability to apply a $\pi$ phase shift rotation on one of the qubits depending on the state of the remaining qubit~\cite{milburn89}. These are trickier to implement than single qubit operations, since this is a nonlinear optical interaction, and such optical nonlinearities, at the single photon level, are extremely weak. An alternative is to mimic nonlinear operations with linear optics and measurement, resulting in a probabilistic gate that provides the correct operation after an appropriate postselection, or with an additional heralding signal.

Historically, a variety of approaches to efficient optical quantum computing were discussed and investigated, for example Ref.~[\onlinecite{gottesman99}] and references therein. However, the field of LOQC took off with the proposal of Knill, Laflamme and Milburn (KLM)~\cite{knill01}, who invented a scalable photonic scheme that required linear optics components, single photon detection and classical feed-forward only (the reader may enjoy reading Ref.~[\onlinecite{rev_myers04}] for comprehensive lecture notes and Ref.~[\onlinecite{rev_kok07}] for a historical overview of KLM). The KLM scheme essentially works by using nonclassical interference to generate a phase shift that is nonlinear with respect to photon number, conditioned on photons appearing at certain heralding modes. These operations are then built into nondeterministic logical gates. The gates are used in a repeat-until-success mode, and the operation of a successful gate is teleported onto the logical qubits. Use of a large number of concatenated steps, and lots of ancilla photons, leads to essentially deterministic gates. The KLM scheme theoretically allowed for a resource-efficient implementation of two- and multi-qubit gates--- unlike encoding a single photon across many modes, the resource scaling was not exponential in the number of qubits, but rather linear. Thus the KLM scheme provides a pathway to build a universal quantum computer, albeit with a large overhead of ancilla qubits (and their associated circuitry) to deal with the use of nondeterministic two-qubit gates. With the advent of a viable theoretical approach, photonic quantum computing became the subject of extensive theoretical and experimental development. As well finding approaches that reduce the overhead due to nondeterminism, making this scheme practical also requires high-quality technological components to make, manipulate and measure~\cite{book_wiseman_2009} the photon qubits. We first turn our attention to these technology considerations. 

\section{Photon Technology}
\subsection{Detecting a photon}

A photon's life in a quantum experiment starts with its generation and concludes with its detection. Both processes need to be efficient, and their performance and properties play essential role in PQC. In this section, we start from the end---with a look at single photon detection~\cite{book_migdall_2013} technology.

An ideal photon detector (PD) clicks every time a photon hits it and immediately restarts its operation. It does not produce false positive signals when no real photons were detected (so-called ``dark counts'') and it also tells exactly how many photons were detected in the same spatio-temporal mode. Such ideal photon detectors do not exist yet. Existing PDs are correspondingly characterized by detection efficiency $\eta_{\rm d}$, reset time $\tau_{\rm R}$ (that sets the maximum detection rate), detection time jitter $\tau_{\rm j}$, dark count rate $C_{\rm d}$, and photon-number-resolving (PRN) capabilities. While a \textit{perfect} PD is not actually required for PQC~\cite{silva05}, improving the PD performance to very high levels is important for a realistic and scalable platform.

Setting aside historical and exotic approaches, the PD of choice for optical quantum information science experiments has been the Si avalanche photodiode (APD) operating in Geiger mode. These are relatively fast ($\tau_{\rm R}\leq100~{\rm ns}$), low-noise (typical $C_{\rm d}\sim100$~counts per second) detectors. Unfortunately, their limited quantum efficiency, typically up to $\eta_{\rm d}\approx65\%$, sets a practical limit on the number of photons that can be used simultaneously in an experiment. A probability of detecting, say, ten photons with ten detectors is already less than $2\%$, and things get exponentially worse with increasing photon number. Si APDs do not possess photon number resolving (PNR) capabilities~\cite{rev_eisaman11} and their maximum efficiency wavelength range is quite limited. In particular, it does not cover the telecommunications bands around $1310$ and $1550~{\rm nm}$. The equivalent detector for $1550~{\rm nm}$, the InGaAs APD, suffers from lower quantum efficiency and higher dark counts.

Inefficient detection was a significant limiting factor for PQC for quite some time. Things  started to turn for the better with the advent of superconducting nanowire single-photon detectors~\cite{goltsman01,rosfjord06} (SNSPDs). These provided something close to a direct substitute for the usual APDs: they have comparable ($\tau_{\rm R}\approx40~{\rm ns}$) reset times, yet can achieve detection efficiencies of up to $\eta_{\rm d}\approx0.93$~(Ref.~[\onlinecite{marsili13}]) (and recently even $\eta_{\rm d}\geq0.95$~(Ref.~[\onlinecite{reddy19cleo}])) in the telecom wavelength range. SNSPDs work by passing a current though a superconducting nanowire close to the critical current---then, the energy absorbed from even a single photon can transition the device to normal resistivity. The subsequent voltage spike is filtered and amplified, and registered as a detection. SNSPDs are a bit more complicated to operate than APDs, as they require cryogenic temperatures of 0.8-3K (depending on the superconducting material), but the massive enhancement in detection efficiency justifies the inconvenience. SNSPD performance can also be optimized to any wavelength by selecting the appropriate material and designing a suitable optical cavity that envelops the nanowire. They can also be designed to efficiently interface with fiber-optic inputs. In short, besides providing an enormous increase in detection efficiency, SNSPDs have enabled operation at telecom wavelength, that benefits from previous development of optical materials and efficient photonic tools. This detector performance is also beneficial for quantum communication and other low-loss applications, e.g.\ Refs.~[\onlinecite{bussieres14,saglamyurek15,shalm15,weston16,valivarthi16,slussarenko17n}].

Research on superconducting detectors is still ongoing, aimed at understanding detection mechanisms in different types of nanowire materials~\cite{renema14,rev_engel15,gaudio16,marsili16,renema17}, improving its performance in terms of reset times~\cite{kerman13}, time jitter~\cite{esmaeilzadeh17,korzh18}, and developing new methods of accurate detection efficiency measurements~\cite{tiedau19}. Although intrinsic dark counts are low, SNSPDs are susceptible to picking up background thermal radiation from the input fiber's room-temperature environment---this can be overcome by spectral filtering.  

The key remaining limitation of this technology is the lack of PNR capability. While schemes that turn SNSPDs into PNR detectors are being investigated~\cite{rev_mattioli15}, a different type of detector, based on transition-edge sensors (TESs)~\cite{cabrera98} can be also employed in experiments where photon number counting is essential. TES detectors work as bolometers with single-photon-level resolution: absorption near the superconducting transition changes the resistance of the device monotonically with photon number, which can be read out through an integrating circuit. TESs have excellent PNR skills~\cite{burenkov17}: in recent experiments they were able to efficiently discriminate up to $\approx20$ photons in the same spatio-temporal mode~\cite{harder16}. At the same time they have shown to be able to reach $\eta_{\rm d}\approx0.95$ in the telecom wavelength range~\cite{lita08}, with further developments leading to even higher $\eta_{\rm d}\approx0.98$~\cite{lita10, fukuda11}, closely approaching the ideal $\eta_{\rm d}=1$. TESs can also be optimized to any wavelength in the visible and IR range. A critical drawback of a TES detector is its slow operation, with $\sim{\rm \mu s}$ reset times and $\geq50 {\rm ns}$ time jitter. Efforts in improving TES time performance are ongoing, with reset times as fast as $\tau_{\rm R}\approx460~{\rm ns}$~\cite{calkins11} and time jitters of down to $\tau_{\rm j}=2.3{\rm ns}$~\cite{lamaslinares13} (for $775~{\rm nm}$ photons) having been demonstrated. Still, these numbers are at least two orders of magnitude higher than might be considered practical for PQC, where clock cycles of $\lesssim 10$~ns are likely required for the practical switching of flying photons.

\subsection{Generating a photon}

Having exceptional detectors isn't much use if one can't efficiently make high-quality photons on which to encode qubits.

Computing tasks in the near and long term require the capability of simultaneously generating a large~\cite{li15} ($N\approx10-10^{11}$) number of single photon states. The obvious way to achieve this is to have a large ($N\approx10-10^{11}$) number of deterministic sources that can simultaneously produce one and only one photon each at the push of a button (i.e.\ on a trigger event). Moreover, these photons must necessarily be: (a) efficiently collected so to be sent into the PQC processor and not lost (e.g. by absorption, scattering, diffraction or mode mismatch during the generation and fiber in-coupling process); (b) in a pure quantum state and indistinguishable from one another; and (c) compatible with the low-loss material and high-efficiency detection technology from above. At present, sources that properly satisfy this list do not exist. However, truly deterministic, high-quality photon sources like this are being developed using diverse physical systems~\cite{rev_eisaman11}, such as trapped ions and atoms, color centers in diamonds, semiconductors, quantum dots~\cite{rev_senellart17}, and other, more exotic, methods (e.g.\ Ref.~[\onlinecite{he17,vogl18,tran17}]). Some of these rely on the use of a single emitter that, in principle, naturally provides on-demand single-photon emission, while others---such as atomic ensemble~\cite{ferguson16} and parametric nonlinear processes~\cite{dellanno06}---require heralding signals and switching to make them so. (The requirements for achieving deterministic operation in practice will be considered in the next subsection.)    

In the meantime, the key enabling technology for experimental quantum optics, spontaneous parametric downconverson~\cite{louisell61,klyshko69,burnham70} (SPDC), remains a practical way to generate high-quality single photons nondeterminstically. Developments in this technology have effectively addressed the feature list (a)-(c) above. In this three-wave mixing nonlinear $\chi^{(2)}$ process, a pump photon from a laser has a small probability to be converted into a pair of `daughter' (signal and idler) photons. The process must obey the momentum ($\vec{k}_{\rm p}=\vec{k}_{\rm s}+\vec{k}_{\rm i}$, phase matching) and energy ($\omega_{\rm p}=\omega_{\rm s}+\omega_{\rm i}$) conservation laws, with $\vec{k}_{n}$ and $\omega_{n}$, ${n}={\rm p},{\rm s},{\rm i}$, being the wavevectors and angular frequencies of pump, signal and idler photons, respectively. SPDC is probabilistic, but it can be used to produce ``heralded'' single photon (and more complex multi-photon~\cite{wagenknecht10,barz10,hamel14,krapick16}) states, where the presence of a photon is heralded by the detection of its twin. Alternatively, SPDC can produce photon pairs that are naturally entangled in polarization~\cite{kwiat95}, transverse spatial modes~\cite{mair01,dada11}, or frequency~\cite{giovanetti02,kuzucu05}. With modest effort, it is possible to produce photon pairs with entanglement in a time-bin encoding~\cite{brendel99}, or even in multiple degrees of freedom simultaneously~\cite{barreiro05}.

SPDC can be a simple and cost-effective way to get single photons and (entangled) photon pairs but, in its original and simplest form, it is far from an ideal photon source for PQC. Ongoing technological development is changing that. Among the immense variety of SPDC-based sources that have been developed and reviewed over past years~\cite{rev_eisaman11,rev_caspani17,rev_ansari18,rev_flamini18}, we concentrate here on some advances that directly serve realistic PQC. 

A typical SPDC output from a simple, critically phase-matched, bulk-crystal source~\cite{kwiat95} is not compatible with efficient coupling into single-mode optical fiber, because its transverse spatial profile is far from a gaussian mode, resulting in coupling loss. This results in coupling loss into single (gaussian) mode fiber. Also, the twin photons are intrinsically entangled in frequency. This means that detection of one photon---to herald the presence of another--- without resolving its wavelength degrades the purity of the heralded photon~\cite{grice97}. The spectral filtering necessary to remove this entanglement adds even more loss to the source. Moreover, traditional SPDC photon wavelengths sit around $800~{\rm nm}$, due to the standard use of Si APDs and compatible with readily-available pump laser wavelengths. At these wavelengths, the material loss (e.g.\ in fibers) is significant, and detection efficiency is limited. 
A typical experiment involving more than one photon pair would have heralding efficiency (probability of a heralded photon to successfully travel from a source to a detector and produce a click~\cite{klyshko80}) of  $\lesssim 10-15\%$, although some experiments report $\approx 30\%$~(Ref.~[\onlinecite{yao12}]). Under these conditions, setting up several photon-pair sources allowed creation of complex photonic states of up to ten photons~\cite{wang16}, but the low collective detection rates, and achievable state quality, limited the long-term prospects of these sources.

A significant step forward was the application of quasi-phase matching~\cite{rev_hum07} (QPM), via periodically poled nonlinear crystals. This expanded the range of possible phasematching wavelengths and emission geometries~\cite{bonfrate99,tanzilli01,sanaka01,banaszek01} and enabled collinear, beam-like downconversion in the telecom wavelength range. With both photons emitted into an almost-single, almost-identical, almost-Gaussian spatial mode, the mode-matching loss and fiber propagation loss could be kept very low, leading to high heralding efficiencies. Using type-II phase matching meant that degenerate photons could be deterministically separated with polarization optics. With the addition of interferometric schemes to generate polarization entanglement~\cite{fedrizzi07,evans10}, QPM SPDC sources could deliver entangled photon pairs with either continuous wave~\cite{kim06,wong06} or pulsed laser pumps~\cite{kuzucu08,scheidl14}. 

There remained the need to remove the residual spectral entanglement in downconverted photons. This was recently solved by applying the concept of group velocity matching (GVM)~\cite{grice97,keller97,konig04}. By carefully engineering the relative group velocities of the pump, signal and idler photons, and adjusting the pump laser bandwidth and SPDC crystal length, the joint spectrum of the daughter photons can be controlled. It can be arranged that the signal photon is in a single spectral mode, and the idler photon is in a single spectral mode, to high fidelity. (Note that the photons do not need to be in the same spectral mode as one another.)  
This technique provides photon pairs that are inherently uncorrelated in their spectrum~\cite{grice01,uren05}, and reduces or removes the need for spectral filtering. GVM at specific donwconversion wavelength sets is attained by selecting an appropriate nonlinear material---KTP (potassium titanyl phosphate) proved to be suitable for degenerate downconversion in the telecom region. Using GVM, a number of frequency uncorrelated~\cite{mosley08,jin13}, non-degenerate~\cite{kaneda16} and degenerate indistinguishable~\cite{evans10,gerrits11,bruno14,greganti18} pure photon-pair sources at telecom wavelength have been demonstrated. Combined with optimized mode matching with the optical fiber~\cite{bennink10} and high efficiency detection technology in telelcom wavelength range, GVM allowed realization of pulsed telecom photon-pair sources that are simultaneously pure, highly efficient and (if desired) entangled in a chosen degree of freedom~\cite{jin14a,shalm15,weston16,slussarenko17n}. Further tailoring of the crystal's nonlinearity profile~\cite{branczyk11,dixon13,graffitti17} provides photons that are fully uncorrelated in their spectrum~\cite{chen17,graffitti18,chen19}, completely removing the need for lossy spectral filtering. Investigation of the performance and limitations of periodically poled SPDC sources continues~\cite{meyerscott17,laudenbach17,zielnicki18,graffitti18a} and even tools for complete SPDC optimization are now available~\cite{spdcalc}. 

These developments have provided an enormous leap forward for SPDC technology, helping it to get close to satisfying many of the criteria (a)-(c) for ideal photon generation. Heralding efficiencies jumped~\cite{slussarenko17n,renema18} 
to above $0.8$. The entangled state quality is harder to survey, because of the variety of figures of merit that are used. Focussing on a couple of standard ones, quantum state purities over $\geq0.997$ (Ref.~[\onlinecite{tischler18p}]) have been observed, and entangled state qualities---equivalent to the fidelity~\footnote{We use $F=[{\rm Tr}\sqrt{\sqrt{\rho}\sigma\sqrt{\rho}}]^2$ as the definition of the fidelity between the states $\rho$ and $\sigma$} with a maximally entangled state---above $0.99$ have been achieved in the lab~\cite{rangarajan09,shalm15,bierhorst18,tischler18p,lohrmann18}. These high-performance sources have also allowed realization of important experiments in entanglement verification ~\cite{giustina15,shalm15} and quantum metrology~\cite{slussarenko17n}. 

However, these advances relate to what happens \textit{when} a photon pair is generated---the pair generation process itself is probabilistic. In the next section, we consider how SPDC or other technologies may be used provide deterministic single photon generation.  

\subsection{Generating a photon deterministically}
\begin{figure}
	\includegraphics{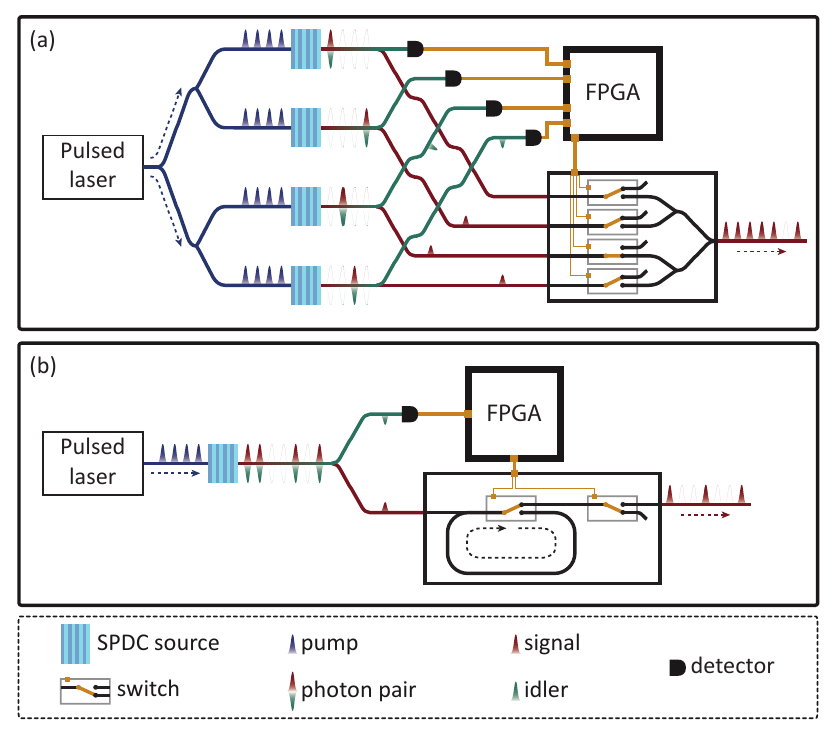}
	\caption{ \label{fig1} Schematic representation of two types of triggered photon sources based on SPDC. (These concepts can be adapted to nondeterministic sources based on other technologies.) (a) Multiplexing scheme that combines multiple (here, four) probabilistic photon sources to provided an increased brightness. Detectors (upper arms) are used to herald the production of a photon by one of the sources, which is then switched into the output mode by some logic (e.g.\ a field-programmable gate array) and switch array. Since there are multiple sources in parallel, this scheme increases such as probability of having a photon in an appropriate time bin, without increasing the probability of having more that one photon in that bin. With enough sources in parallel and with low loss, the arrangement can approach an ideal, deterministic single photon source. (b) Triggered source that uses only one probabilistic source and an active delay network. The network rearranges the generated photons in time, so that they are output at a stable, although lower, repetition rate. This scheme also provides a deterministic source, in principle.}
\end{figure}

Photon-pair sources from SPDC and related processes---like spontaneous four-wave mixing (SFWM)~\cite{fiorentino02,rarity05,fulconis05,fan05}---are not only nondeterministic but generally operate at low generation probabilities. In order to keep the single photon state quality high, pump powers have to be kept low, otherwise multiple photon pairs will be generated at the same time~\cite{fulconis07}. This limits practical photon-pair generation probability $\xi$, for SPDC and similar processes, to $\xi \lesssim 1\%$. Directly combining an array of $n$ such sources (that will together produce $n$ simultaneous pairs with probability $\xi^n$) to generate a larger quantum state is essentially not a viable option for a scalable photonic quantum computer.

A more feasible alternative is to employ a deterministic photon source. In recent years, photon-on-demand sources based on quantum dots~\cite{rev_gazzano16}, both free-space~\cite{gazzano13,somaschi16,rev_senellart17} and integrated~\cite{kim17,aghaeimeibodi18,dutta18} into optical waveguides, have demonstrated a significant increase in brightness, enabling new quantum computation experimental demonstrations~\cite{he17b}. (It is worth noting that although quantum dots are usually assumed to provide single photons on demand, quantum dots can also generate entangled photon pairs~\cite{dousse10,bennett10,jayakumar14,heinze17,prilmuller18,huber18} and superpositions of photon number states~\cite{loredo19}.) Although quantum dots~\footnote{We have focused on quantum dots as these are the leading contender to SPDC/SFWM for photon sources in PQC. However, as mentioned earlier, a variety of other single emitters are also suitable in principle.} can couple to optical cavities with very high efficiency~\cite{englund07,somaschi16,wang19qd}, a currently outstanding problem is coupling light efficiently into single mode optical fibers, with present coupling efficiencies~\footnote{Here, we mean the ratio of the rate of photons coupled to the rate of trigger pulses.} $\lesssim 33\%$~(Ref.~[\onlinecite{wang17b}]). Moreover, each quantum dot is usually spectrally different from others due to structural and environmental inhomogeneities, so the photons emitted by two dots are distinguishable from each other. PQC relies on non-classical interference, and the lack of indistinguishability makes it complicated to increase the number of photons used simultaneously in an experiment. One way to fix this is to  tune the emission spectrum of different quantum dots to make indistinguishable ~\cite{patel10,ellis18}. Alternatively, a single quantum dot can be used to generate all the required photons. For this, a pulsed output stream of photons from the dot is demultiplexed into different spatial channels via a free-space~\cite{wang18a,anton19} or integrated~\cite{lenzini17} active optical network. The multiplexed photons are then each delayed by appropriate amounts, so as to be output simultaneously from the source setup. 

Similar active optical circuits can also, in principle, turn probabilistic sources such as SPDC into deterministic ones. To realize this, an array of sources is used---see Fig.~\ref{fig1}(a). Detecting the heralding signals from such an array will label which source has successfully generated a photon pair. Then, the corresponding heralded photon can be actively re-routed through an optical network towards the output, while other photons, if generated, would be discarded by the same network. Using $n$ sources this way theoretically boosts the generation efficiency to $\xi_\textrm{multi} = 1-(1-\xi)^n$, ideally, without increasing the pump power that impinges on a single nonlinear crystal and thus without increasing the amount of high-photon-number noise from multiple-pair generation events. (In principle, the network can also filter out multiple-pair generation events if photon-number resolving detectors are used.) This concept~\cite{migdall02,shapiro07}, experimentally demonstrated in 2011~(Ref.~[\onlinecite{ma11}]), has moved significantly towards practicality since then~\cite{collins13,meany14,francisjones16}, in part because of the use of fiber- and waveguide-based integrated platforms to help scaling.

Another method, that does not require multiple separate sources, is to use time~\cite{pittman02,migdall02,jeffrey04} (or frequency~\cite{grimaupuigibert17,joshi18})  multiplexing of a single source~\cite{pittman02,migdall02,jeffrey04}. In the time multiplexing approach, shown in Fig.~\ref{fig1}(b), a heralded photon pair is generated in a random time bin, but the timing is recorded through detection of the heralding signal. The heralded photons are sent into an active temporal delay network and switched so as to exit the network at a fixed, although lower, repetition rate. The number $n$ of time bins that is used to output one single photon plays the role of $n$ sources in a spatial multiplexing scheme. Thus the improvement in generation probability scales with the size of the delay network, but is affected negatively by the loss in optical components in it. This multiplexing idea has been recently implemented in a number of experiments, demonstrating multiplexing with large-scale~\cite{mendoza16}, or large-scale and low-loss~\cite{kaneda15} networks, or with devices that produce indistinguishable output photons~\cite{xiong16}. The experimental demonstration that includes all of these features~\cite{kaneda19} produced single photons in the output fiber with a probability of $\approx 0.6$, and these photons displayed a non-classical interference visibility $\approx0.9$. A more in-depth look at near-deterministic sources can be found in Ref.~[\onlinecite{rev_caspani17}].

Interesting preliminary work has also been done towards combining these kinds of techniques to simultaneously generate more than one single photon at a time. The multiplexing approach can be applied to more than one probabilistic source to generate states with one photon in each of $N>1$ modes~\cite{gimenosegovia17,zhang17}. An alternative method is to use an optical quantum memory to synchronize several probabilistic sources~\cite{kaneda17}. Although quantum memory might be as simple as a switchable optical delay (in a free-space, fiber, or waveguide loop, for example), there is also extensive theoretical and experimental development of memories based on matter systems~\cite{rev_lvovsky09}, with recent achievements including but not limited to broadband~\cite{saglamyurek11,saunders16}, high-speed~\cite{kaczmarek18}, multimode~\cite{ferguson16,tiranov16}, telecom-compatible~\cite{saglamyurek15,rancic17}, or configurable~\cite{campbell14} memories, capable of storing vector-~\cite{parigi15}, vortex-~\cite{yang18}, or entangled-~\cite{saglamyurek15,tiranov16} qubits, and storage with long coherence times~\cite{zhong15}, high storage efficiency~\cite{hedges10,cho16,wang19} and fidelity~\cite{hosseini11,vernazgris18}.\blk

Over the span of slightly more than a decade, photon detection and the probabilistic generation of high-quality photons have undergone transformational advances, and the development of deterministic sources is well underway, with no in-principle barriers to their realization. (There are also other interesting advances, such as spectrally narrowband sources~\cite{rambach16} for metrology and fundamental physics applications~\cite{rambach18}, that we do not cover here.) 

\subsection{Manipulating a photon}
Thus, before proceeding to the next section, we briefly turn our attention to technologies for manipulating photons for PQC. Precise and accurate control of photon's polarization, path or time-bin stat has always been the strength of PQC~\cite{ralph09}. Recently, this has been extended to performing reconfigurable mode transformations in integrated quantum optics~\cite{wang18s}. Modern electro-optic elements, such as Pockels cells or integrated electro-optic modulators, allow fast polarization switching sufficient perform rigorous Bell tests with locality and freedom of choice loopholes closed~\cite{giustina15,shalm15} or spatial mode switching for source multiplexing purposes~\cite{lenzini17,wang18a,anton19}. Efficient tools for manipulation of more exotic degrees of freedom, such as frequency-time~\cite{eckstein11a}, or transverse spatial modes~\cite{slussarenko13} are also being developed, including the techniques that transfer information from one degree of freedom to another, such as polarization to spatial transverse mode~\cite{dambrosio12}, discrete variable to continuous variable~\cite{sychev18}, frequency conversion~\cite{kasture16}, and so on.

\subsection{Integrated quantum photonics}
\begin{figure*}
	\includegraphics{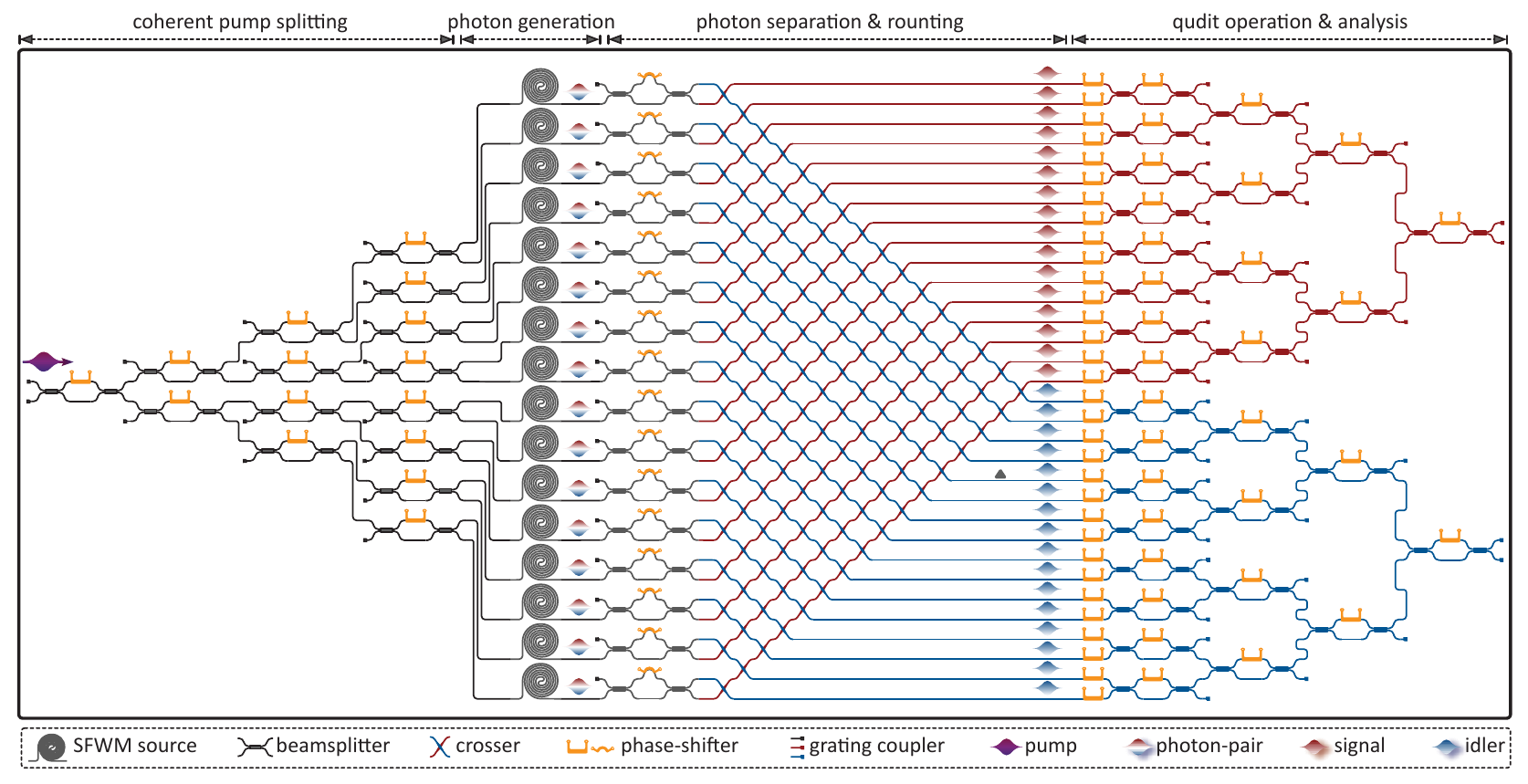}
	\caption{ \label{fig2} A circuit diagram of the multidimensional silicon quantum photonic circuit. Reprinted with permission from Wang et al., Science 360, 285–291 (2018). Copyright 2018 AAAS. The device monolithically integrates 16   photon-pair sources, 93 thermo-optical phase shifters, 122 multimode interferometer beamsplitters, 256 waveguide crossers, and 64 optical grating couplers. A photon pair is generated by SFWM in superposition across 16 optical modes, producing a tunable multidimensional bipartite entangled state. The two photons, signal	and idler, are separated by an array of asymmetric Mach-Zehnder-Interferometer (MZI) filters and routed by a	 network of crossers, allowing the local manipulation of the state by linear optical circuits. Triangular networks of MZIs perform arbitrary local projective measurements. The photons are coupled off the chip into fibers by means of grating couplers, and are detected by two SNSPDs. See Ref.~[\onlinecite{wang18s}] for details.}
\end{figure*}

While introducing the relevant advances in photon  detection and generation technology, we mostly limited ourselves to the ``bulk'' optics environment, with separate optical components sitting on a tabletop.  
As the scale of PQC demonstrations grows to larger numbers of photons and gates, the importance of technological scalability and miniaturization becomes increasingly apparent. Integrated waveguides and optical chips offer an obvious path to implementing circuits at scale, i.e.\ with huge numbers of components packaged compactly. Thus, these technologies are now playing a significant role in the field~\cite{rev_tanzilli12}. Several characteristics are important for a waveguide platform: the achievable density of optical components; low propagation and coupling losses; and the ability and speed of active reconfiguration, for example. It is also desirable to integrate sources and detectors onto the optical chip. 

Different materials offer their own strengths and advantages for realizing a practical integrated quantum photonics platform. Femtosecond-laser-written waveguides (typically in a glass) support polarization qubits~\cite{sansoni10} and are not restricted to a 2D geometry, allowing realization of complex couplings in 3D interferometric networks~\cite{crespi16}. Lithium niobate, a material that is already well established in classical integrated photonics, is an efficient and flexible platform for photon sources and fast switchable electro-optical components operating at the GHz rates. Both ion-indiffused and high-index-contrast etched waveguides are being developed and employed~\cite{bonneau12,lenzini17,hopker17,krasnokutska18,krasnokutska19,aghaeimeibodi18}. Silicon-based optical chips offer high component density, low loss, the ability or potential to integrate every necessary component, and compatibility with existing foundry processes~\cite{silverstone16}. An enormous range of other materials platforms are also under consideration.

On the integrated detection front, a lot of work has been done~\cite{rev_ferrari19} in embedding SNSPDs into optical chips since the first demonstration in 2011~(Ref.~[\onlinecite{sprengers11}]). This ongoing effort has already provided fast and efficient~\cite{pernice12}, low-noise~\cite{schuk13}, fast and low-noise~\cite{kahl15}, or low-noise, efficient and  fast~\cite{akhlaghi15} (and even faster~\cite{munzberg18}) detection at telecom wavelength. Significant effort is being put into turning waveguided SNSPDs into waveguided PNR detectors, see for example Ref.~[\onlinecite{hopker17}] and references therein, and Ref.~[\onlinecite{rev_mattioli15}]. Similar developments are happening on the TES integration side~\cite{calkins13,hopker19}.

The situation is even more vivid regarding integrated photon sources. QPM-based downconversion, which now plays the key role in heralded photon and photon-pair generation, was in fact first demonstrated in fiber~\cite{bonfrate99} and integrated waveguides~\cite{tanzilli01,sanaka01,banaszek01}. An important advantage here is the transverse spatial confinement of the three (pump, signal, idler) propagating optical modes along the entire length of the nonlinear material. This confinement allows construction of a photon-pair source with both high brightness (absolute generation rate calculated in pairs per second per mode per unit of pump power) together with high heralding efficiency. This is advantageous compared to bulk SPDC, where the spatial mode configuration for high brightness is different from the one that provides high heralding efficiency~\cite{bennink10}. Using integrated technology, efficient sources in the telecom wavelength range~\cite{zhong10}, including ones with GVM~\cite{eckstein11,harder13}, have also been realized, leading to the development of fully-packaged,  banana-sized~\cite{montaut17},  and highly efficient photon-pair source; and similar sources in a variety of material platforms~\cite{atzeni18,meyerscott18}. Techniques have been demonstrated for direct and practical characterization of nonlinear operations (like SPDC) in integrated quantum photonics~\cite{lenzini18}.
Integrated optics has also shown the capability of using more than one degree of freedom of a photon~\cite{atzeni18}. 

A number of materials for integrated optical components have no $\chi^{(2)}$ nonlinearity, making them unsuitable for SPDC-based photon-pair sources. In this case, a practical alternative is SFWM. It is a $\chi^{(3)}$ nonlinear parametric process where two pump photons (degenerate or otherwise) are converted into two daughter photons (degenerate or otherwise), conserving energy and momentum. Historically investigated in optical fibers~\cite{fiorentino02,rarity05,fulconis05,fan05} due to the isotropic nature of amorphous silica, this method is now commonly adopted in those integrated platforms where $\chi^{(3)}$ nonlinearities dominate~\cite{sharping06,davanco12}. A GVM-like approach for controlling the joint spectrum of daughter photons was also subsequently generalized to SFWM~\cite{garaypalmett07} and implemented experimentally in fiber~\cite{smith09} and on a chip~\cite{spring13}. The scalability of the integrated optics approach allows one to fabricate arrays of nearly identical photon sources~\cite{silverstone14,spring17} that are now actively used in PQC experiments in silicon~\cite{wang18s}.
On the more technical side, a number of SFWM obstacles, including the challenge of strongly filtering out the strong pump field from the generated photon field, have been overcome in recent years~\cite{harris14,grassani16}. The interested reader can find more information on integrated probabilistic sources in Ref.~[\onlinecite{rev_caspani17}] and on recent advances in GVM bulk and waveguided sources in Ref.~[\onlinecite{rev_ansari18}].

The rapid development of quantum integrated photonics is perhaps most obvious in the growth in the scale, complexity and performance of optical circuits for one- and multi-qubit operations. The first optical chips with path-~\cite{politi08} and polarization-~\cite{sansoni10,crespi11} qubit encoding did not immedaitely surpass the performance previously achieved with bulk optics~\cite{lu07,lanyon07} (in repeating the factoring of $15$ by a compiled Shor's algorithm, for example~\cite{politi09}), but emphatically demonstrated the promise of the integrated approach. Subsequent devices, and the applications they implemented, started to increase in complexity really quickly~\cite{shadbolt11}. This included increasing the number of interferometers on a chip (Fig.~\ref{fig2}) and adding slow or fast active phase~\cite{matthews09,smith09a,bonneau12,flamini15} and spectral~\cite{notaros17,krasnokutska19} controls in various waveguide platforms. These capabilities have led to a realization of fully-reconfigurable optical processors for an increasing number of optical modes~\cite{carolan15}. It has been observed that, for the moment at least, the number of components on integrated quantum photonics chips is undergoing a Moore's-law-like exponential growth with time~\footnote{M.~G. Thompson, presentation at QCrypt 2016, see \url{http://2016.qcrypt.net/wp-content/uploads/2015/11/Invited3_Mark-Thompson.pdf} (accessed 11 June 2019)}.

A challenge of integrated platforms is optical loss caused by material absorption, waveguide roughness, and coupling onto and off chip. These are actively investigated by a variety of techniques including: improved materials (e.g. higher purity); moving to high-index-contrast platforms where devices can be smaller (e.g.~Ref. \onlinecite{krasnokutska18}), by integrating sources and detectors directly on chip. Modular architectures are also being investigated~\cite{mennea18}.

\section{Quantum computing}

The advent of the KLM scheme~\cite{knill01} in 2001, with its proof of the scalability of optical processing, inspired a worldwide push towards a universal quantum computer with photons. Of course, a full-scale error-corrected version could not be built at that time, and indeed universal quantum computer remains a challenging quest today in any quantum system. The KLM scheme led to the development and improvement of a variety of photonic encodings, schemes for quantum gates, and protocol and algorithm demonstrations~\cite{rev_ralph09,rev_kok07,rev_obrien09,rev_ladd10}. Circuit-based approaches, having evolved from KLM, continue to be an active area of theoretical and experimental research as a path towards intermediate-scale and universal quantum processors.

A significant development for PQC was the realization that the cluster-state model of quantum computing (also known as one-way quantum computing)~\cite{raussendorf01} was well-suited to photon qubits\cite{nielsen04}. This is primarily because large cluster states~\footnote{Cluster states are graphs where the nodes are qubits and the edges are entangling links between the qubits, satisfying particular constraints~\cite{raussendorf01}.} can, in principle, be built efficiently using entangled photon sources and teleportation gates of the kind used in the KLM scheme. It is also important that photon measurements are easy to perform reliably, and because cluster state schemes can be made tolerant to photon loss, the primary source of noise in an optical environment. For these reasons, cluster state schemes are widely viewed as offering a realistic path to scalable PQC.    

As the development of universal PQC has continued, a number of intermediate goals have emerged, providing short- to medium-term targets and a path towards full-scale devices. These include: the development of individual quantum gates of increasing complexity in the circuit model; the implementation small-scale quantum algorithms and non-universal circuits or clusters for them; the development of simplifying and supporting techniques within the circuit and cluster-state models; and the advent of algorithms for sampling problems based on the fundamental properties of bosons. Intermediate quantum computing research is helpful for optimization of the general schemes for PQC, and for developing and testing of individual components of a future quantum computer. 

\subsection{Intermediate quantum computing}
\begin{figure*}
	\includegraphics{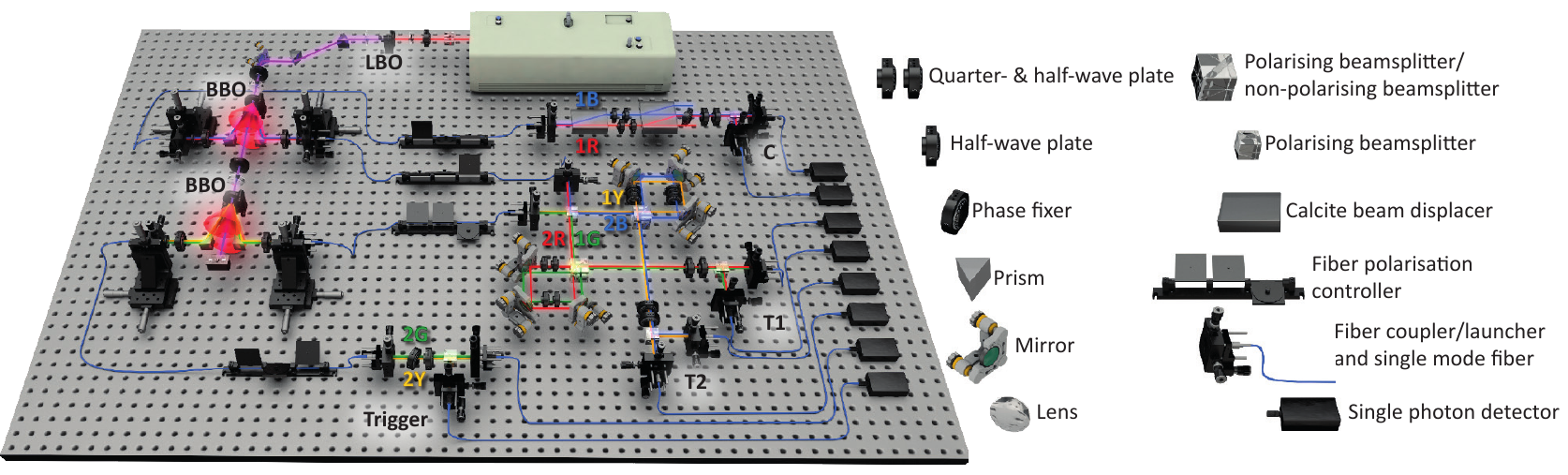}
	\caption{ \label{fig3} An optical quantum Fredkin. Reprinted with permission from Patel et al., Sci. Adv. 2, e1501531 (2016). Copyright 2016 Author(s), licensed under a Creative Commons Attribution 4.0 NonCommercial License. The Fredkin (or controlled-SWAP) gate uses the method of adding control to an arbitrary untiary operation. Entangled photons are produced in BBO (beta-Barium borate) crystals via SPDC. The control qubit is encoded into modes 1B and 1R, target 1 is encoded on modes 2R and 2B, and target 2 is encoded on modes 1G and 1Y. The control circuit consists of a polarization beam displacer interferometer. The path-entangled state, required for the Fredkin operation, is produced after each target photon enters a displaced Sagnac interferometer and the which-path information is erased on a non-polarizing beamsplitter. Quarter-wave plates and half-wave plates encode the target qubits' input state. Successful operation is heralded by fourfold coincidence events between the control, target, and trigger detectors. See Ref.~[\onlinecite{patel16}] for details.}
\end{figure*}

Photons can be readily and accurately manipulated at the single-qubit level---very high fidelity one-qubit gates can be constructed\cite{peters03} because of excellent optical mode control\cite{carolan15}. Initially, particular attention fell on the controlled-NOT (CNOT) gate to complete a universal gate set in the quantum circuit model.  
Theoretical proposals for nondeterministic CNOT gates~\cite{ralph02,pittman01}, demonstrating the basic measurement-induced nonlinearity concept of KLM, were quickly followed by experimental CNOT demonstrations~\cite{obrien03,pittman03} and characterizations~\cite{obrien04,rohde05}. These were expanded to include heralded KLM-style~\cite{okamoto10} and teleportation~\cite{bennett93}-based~\cite{gottesman99} schemes~\cite{gao10a}. A number of proof of principle algorithms followed the early demonstrations of photonic gates~\cite{rev_ralph09}. While using CNOTs to build arbitrary unitary circuits is, of course, a working theoretical method, it is far from optimal. This is because, for example, the decomposition of a three-qubit gate, such as the Toffoli gate, into one- and two-qubit operations may require a large number of such gates~\cite{nielsen11}.  An alternative would be to look for ways of implementing gates that can operate on a larger number of qubits directly.

An interesting and important class of arbitrary-scale quantum logic is the family of controlled-Unitary (CU) gates. In these, a (possibly multi-qubit) unitary operation acts or not---depending on the state of a control qubit---on the target qubits. CU gates are important in various computational tasks, for example the phase estimation algorithm that underlies Shor's algorithm~\cite{lanyon07,lu07} and in quantum chemistry~\cite{lanyon10,santagati18}. A key realization is that implementing the unitary operation $U$ alone may be possible or even easy, but adding the control operation---i.e. conditional action---is difficult.  

A general scheme for adding a control operation to an arbitrary unitary transformation was proposed in 2009~(Ref.~[\onlinecite{lanyon09}]). In this method, given the unitary to be controlled, the Hilbert space dimensionality of the incoming target qubits is first doubled by using some auxiliary degree of freedom of the corresponding photons. Half of the modes of each target qubit pass through the unitary, while the remaining half bypass it. Then, the control qubit state is used to route the target qubits to either pass the unitary or bypass it, via the corresponding modes. After that, the modes are recombined, so the Hilbert space is shrunk to its original dimensionality. This effectively creates a CU gate. (The scheme can be simplified even further, by substituting Hilbert-space-expanding gates with photon sources that generate entanglement in the auxiliary degree of freedom. The term ``entanglement-based'' is usually used in the literature to describe these types of gates, which are not completely general due to the need to generate the initial entanglement, but can be useful at circuit inputs.) This overall method is particularly suitable for optical quantum computing, because high dimensional systems, multiple degrees of freedom, and means of transfering information between them are readily available. Moreover, theoretical studies also highlighted that adding control to arbitrary unitary gates is generally impossible for matter-based qubits~\cite{araujo14,thompson18}, so the method demonstrates a benefit of using fields to quantum compute.

This general approach was used to experimentally realize arbitrary controlled-single-qubit unitaries, a CNOT gate~\cite{zhou11}, and three-qubit gates---namely the Toffoli~\cite{lanyon09} and Fredkin (controlled-SWAP, see Fig.~\ref{fig3})~\cite{patel16} gates. It was also employed in experimentally implementing a number of quantum computing tasks, such as solving systems of two linear equations~\cite{cai13} (this was also done without entanglement-based gates~\cite{barz14}), factoring $21$ by a version of Shor's algorithm (Ref.~[\onlinecite{martinlopez12}]), measuring state overlaps and state purity~\cite{patel16}, and eigenstate witnessing for simple quantum algorithms~\cite{santagati18}.
Entanglement-based gates are now also used in larger quantum circuits, including the ones realized in an integrated platform~\cite{santagati18,qiang18}.

The use of various photonic quantum gate architectures has allowed realization of a variety of intermediate scale simulations, implemented in bulk and integrated optics platforms. Among these~\footnote{Earlier reviews provide references to samples of older demonstrations in the field.} are spin chain simulation~\cite{pitsios17}, calculating molecular ground-state energies\cite{peruzzo14}, Hamiltonian learning~\cite{wang17} and eigenstates witnessing~\cite{santagati18}, and complex state transformations such as 
Fourier~\cite{weimann16,crespi16} or Kravchuk~\cite{stobinska18} transforms.

A highly topical intermediate photonic quantum computing task is that of BosonSampling~\cite{aaronson13,broome13,spring13s,tillmann13,spagnolo14,carolan14,zhong18,rev_brod19}, which is an example of sampling-type computational problems more generally~\cite{lund17}. BosonSampling is a non-universal protocol for which there is strong theoretical evidence that a quantum advantage can be observed. Consider $n$ single photons input into $m \gg n$ optical modes, which are subjected to a random unitary operation on the mode space. It is classically computationally hard to obtain samples from the probability distribution representing where the photons appear at the output. By contrast, photons (and other bosons) traversing a unitary on the mode space perform this calculation naturally. Interestingly, the same quantum-classical performance divide exists even if the photons are allowed to arrive at random inputs of the circuit~\cite{lund14}.  It is thought that better-than-classical BosonSampling performance may be achieved with 50-100 photons, promoting the idea that this system could well provide the first rigorous experimental demonstration of a quantum computational advantage. Nevertheless, challenging constraints on photon loss and other noise still need to be met to achieve this goal~\cite{rahimikeshari16,neville17,renema18}. Recent reviews~\cite{rev_flamini18,lund17} cover conceptual and experimental aspects of the topic in more detail. 

Intermediate quantum computing is likely to lack fully-fledged error correction. Thus, photon loss and noise in PQC will need to be controlled by other methods. One prominent approach being investigated for NISQ~\cite{preskill18} (noisy intermediate-scale quantum devices) is machine learning (ML). ML provides a method to work with quantum protocols operating in an environment of unknown or uncharacterised noise, or where the full \textit{ab initio} modelling of the protocol is intractable~\cite{niu19}, and can be applied to PQC and other systems~\cite{mavadia17}. The flip side to ML helping quantum computation by controlling noise is the hope that quantum computers can enhance ML for other applications, possibly even in the NISQ regime~\cite{preskill18}. Other relevant proposals for ML quantum applications include long-distance quantum communication~\cite{wallnofer19} or metrology~\cite{hentschel11,lumino18}. Experimental demonstrations of ML application to quantum information science have recently started to appear, too~\cite{cai15,wang17,gao18,pepper19}.

\subsection{Cluster-state based computing}
In conventional PQC, uncorrelated input qubits are processed by a complicated quantum circuit of one-, two-, three-, and many-qubit gates (which in turn can be decomposed to one- and two-qubit gates). Here, generating many uncorrelated photonic qubits is considered the ``easy'' part of the problem, and the logical circuit does the ``hard''~\footnote{The terms ``hard'' and ``easy'' are rather qualitative, but they give an idea of the motivation for this approach.} task of performing the  computation. An alternative approach is one-way (or cluster-state) quantum computing~\cite{raussendorf01,raussendorf03,nielsen04}. In one-way computing, a hard-to-make, highly-entangled multi-photon state is sent into an easy-to-implement processing circuit that consists only of single-qubit operations, measurements, and classical feed-forward~\cite{rev_briegel09,rev_ralph09}. The key idea is that, in the absence of deterministic two-photon operations, the cluster state can be built up offline using nondeterministic interactions, and then the computation progresses via those deterministic single-qubit operations for which optics is especially suited. 

Follow-up development showed how to create cluster states more efficiently~\cite{browne05}, leading to significantly reduced resource requirements (characterized as Bell-pairs per effective two-qubit gate, a metric of PQC overhead; smaller $\leftrightarrow$ better) compared to many other optical schemes. 
The one-way computing approach is also more tolerant to losses, compared to KLM~\cite{varnava08}. Since the first experimental demonstration of the essentials of one-way quantum computing~\cite{walther05}, considerable steps have been made towards making larger cluster states~\cite{lu07n,zhong18}, demonstrating larger computing networks~\cite{greganti16}, and improving the feed-forward performance~\cite{prevedel07}. A type of one-way-based computing, where the computer cannot determine the input data and performs the computation blindly but correctly, has also been demonstrated~\cite{barz12}. Developments in the theory of optical one-way computing have driven increasingly realistic schemes for large-scale photonic quantum computing.

Indeed, recent theoretical developments suggest that cluster-based quantum computation may be a more realistic approach towards the future photonic quantum computer than gate-based models. 
There are a number of key advantages to a cluster-state approach. One concerns the way that clusters are built, through progressive nondeterministic fusion operations~\cite{browne05,ewert14} that seek to merge two smaller entangled states into a larger one. The key point is that the failure of the nondeterministic operation slightly reduces the sizes of the initial entangled states, but does not destroy them~\cite{browne05}. In fact, it has been shown theoretically that missing links and nodes (e.g. due to fusion failures or optical loss) in the constructed cluster state need not be problematic. As long as their prevalence is below a certain threshold, percolation theory can be used to reshape the entangled state and perform universal computation~\cite{kieling07,pant19}. The percolation operation corresponds, roughly, to a classically-efficient relabeling of the cluster. Furthermore, error correction for fault-tolerant quantum operations seems achievable, especially given modest loss thresholds~\cite{varnava06,varnava08,rudolph17,morley-short19}. 

In principle, cluster states can be generated and processed (via adaptive measurement) on the fly, without the need to store photons in an optical quantum memory. This is known as ballistic cluster state computing~\cite{gimenosegovia15}. In this scheme, an array of sources and simple circuits produce entangled photons at each time step - these photons are entangled together to produce a 3D cluster where each layer represents a generation step in time. It has been shown that the depth of cluster that needs to exist at any time is only of the order of a few tens of photons~\cite{morley-short17}. In this case, given a suitably small number of faults in the cluster, the computation can proceed indefinitely in principle, with the source array continuing to make new cluster layers at each time step and detectors measuring a layer at each step.

Ongoing theoretical and experimental research on photonic clusters and ballistic schemes is also addressing many technical details (e.g.\ optimal cluster geometry, error correction schemes, sources designed for cluster generation). However, it is emerging that photonic cluster schemes, and closely related ideas, are extremely plausible approaches for realising universal quantum computers~\cite{rudolph17}. 

In summary, in the span of less than two decades, photonic quantum information science has matured immensely. New photon generation and detection technologies have enormously improved the efficiency and quality of photonic quantum states. Integrated circuits grew from a simple demonstration of a beam-splitter to massively-multimode reconfigurable circuits. The number of photons simultaneously used in experiments has grown, from 2-4 up to 12~(Ref.~[\onlinecite{zhong18}]). Overall, experimental PQC is steadily moving towards the major goal of universal quantum computing and theoretical PQC is steadily progressing towards more resource-efficient and noise-tolerant schemes. In parallel, non-universal quantum computation schemes such as BosonSampling are also rapidly scaling up towards the demonstration of the true quantum computational advantage over classical computers.

\section{Networking quantum processors}
 PQC is strongly interlinked with other optical quantum information tasks. On one hand, quantum phase estimation algorithms, used in e.g.\ Shor's algorithm and a number of intermediate quantum computing schemes (as in Ref.~[\onlinecite{peruzzo14}]), are also useful in quantum-enhanced metrology~\cite{wiseman09,higgins07,berni15,daryanoosh18}. On the other hand, quantum communication is essential for building a distributed quantum processor from interlinked quantum computers. Flying fast, photons (or other optical states) are the obvious way to transmit quantum information. Thus photonic quantum interconnects can naturally be tasked with interfacing remote systems and, perhaps, local processing cores. Optical connections make sense regardless of the quantum system chosen for processing, but using photonic processing means that the interconversion between a stationary and a flying qubit can be skipped. (Indeed, quantum teleportation---an entanglement-based protocol used in communication---also plays a key role in a number of PQC approaches~\cite{gottesman99,knill01}.) Nevertheless, it may be that there is some need to adjust the spectral properties of photons between the communication and the processor, and ways to do this are being investigated for a variety of different interconversion wavelengths, and photon-carrying and generating architectures~\cite{rakher10,ikuta11,degreve12,bell16,rutz16,kasture16,dreau18}.

Creating verified communication links capable of sharing and transmitting entanglement is essential for networking quantum computers, and also quantum secure communication, small communication-based processing tasks (quantum communication complexity~\cite{buhrman10,wei19}), and quantum networks for distributed metrology~\cite{gottesman12}. A major step in entanglement verification and distribution was the experimental implementation of loophole-free Bell tests, executed with photonic~\cite{giustina15,shalm15} and matter qubits~\cite{hensen15}. Besides definitively showing local realistic explanations of entanglement are not viable, these tests confirmed that entanglement can now be rigorously verified in a loophole-free manner, opening the road to the unconditionally-secure device-independent protocols~(e.g.\ Ref.~[\onlinecite{acin07}]).  A remaining challenge is enabling these protocols in the presence of very high loss in a communication channel used to distribute the entanglement. As in PQC, loss is the predominant source of added noise that degrades entanglement. 

One can neglect the loss by postselecting only on successful detection events, however such experiments do not offer device-independent security or a quantum advantage in metrology. Unfortunately, the no-cloning theorem forbids creation of identical backup copies of unknown quantum states to be used if a photon is lost. A state-independent attempt to amplify a qubit or qudit (i.e. to boost the photon number to its original value) would inevitably lead to the degradation of the state purity. Noiseless amplification can only be performed in probabilistic manner---consistent with noise reduction being a non-unitary process---and produces a wrong output upon failure. Fortunately, heralded amplification (also known as noiseless linear amplification, NLA) is possible: in this probabilistic scheme, successful amplification events are heralded by an independent photon detection signal, allowing them to be sorted from the failed trials~\cite{ralph09}. Heralded amplification can be used to distribute entanglement in the presence of loss---even with the detection loophole closed, in principle. Since the first demonstrations~\cite{xiang10,ferreyrol10,zavatta11}, NLA has been actively researched in both the discrete-variable (photon) and continuous-variables communities. It has shown the ability to amplify polarization~\cite{kocsis13}, path~\cite{monteiro17} and time-bin~\cite{bruno16} qubits, and has been used to restore mode entanglement that was degraded due to loss~\cite{xiang10,ulanov15,monteiro17} and versions of the scheme have been applied to quantum communication~\cite{chrzanowski14} and cloning~\cite{haw16}. There have been proposals and experiments related to implementing NLAs with quantum logic gates~\cite{mcmahon14,ho16}.

Many other communication protocols for sharing high-quality entanglement in lossy environments (i.e. for realising quantum repeaters) are based on entanglement swapping~\cite{jennewein02}. Recent advances have used entanglement swapping for sharing entanglement with the detection loophole closed, even over high-loss channels~\cite{hensen15,weston18}. Other potential tools include quantum nondemoliton measurements of photon number~\cite{pryde04,meyerscott16} and, of course, a variety of error-correction-code protocols (e.g.~Ref.~[\onlinecite{bolt16}]). Ultimately, entanglement-based networks will likely also require local processing (i.e.\ small quantum computers) for distilling entanglement, and quantum memory for synchronizing operations.

\section{Conclusion}
Our short review has only touched briefly on other PQC elements including error correction in photonic schemes~\cite{auger18}, optical quantum memories, and algorithms and protocols. There is also a broad range of related research that is beyond our immediate scope, including other qubit or qudit encodings---such as single-rail~\cite{lund02,rev_ralph09}, parity state~\cite{gilchrist07,rev_ralph09}, continuous-variable~\cite{menicucci06,menicucci14,yokohama13,chen14}, and hybrid~\cite{sychev18,drahi19}---as well as other source and detector technologies. Some of these techniques are also promising in terms of resource use and scalability. Instead, we have covered technologies and methods that are the main focus of the experimental development of photonic (Fock-state) quantum information processing in the medium term, and provide a firm foundation for the development of large-scale devices. 

There is significant promise for the long term. Improvements in cluster-state schemes designed specifically for photonics are providing a reduction in the overhead (from nondeterminism) and in error thresholds---especially for loss. In conjunction, exceptional quality sources, detectors and gates---and large-scale integrated platforms---are providing the hardware advances required to build processors comprising very many elements. Intermediate tasks like BosonSampling provide a path to demonstrating a true quantum computing advantage sooner rather than later. And photonics continues to be the dominant platform for connecting processors separated by distance, and for remote entanglement sharing in general. 

There remain other potentially transformational technologies for photonic processing. We have only touched briefly on nonlinear interactions at the single-photon level---mediated by atoms, for example. Such schemes, applied at scale, could massively reduce the overhead of ``linear plus measurement'' approaches. However, there remains significant research and development required to capitalize on their promise. In the meantime, or perhaps in their stead, the convergence of technological performance and theoretical requirements in photonic linear optics is pointing to a bright future for photon processing.

\noindent \textbf{Acknowledgements.} This work was conducted by the ARC Centre of Excellence for Quantum Computation and Communication Technology under grant CE170100012. 

%

\end{document}